\documentclass[a4paper,fleqn,usenatbib]{mnras}

\usepackage[T1]{fontenc}
\usepackage{ae,aecompl}

\pdfoutput=1

\usepackage{graphicx}	% Including figure files
\usepackage{amsmath}	% Advanced maths commands
\usepackage{amssymb}	% Extra maths symbols
\usepackage{geometry}
\usepackage{pdflscape}  % landscape pages
\usepackage{dcolumn}    % aligning numbers by decimal points in table columns
\usepackage{upgreek}    % for non-italic \mu
\usepackage{soul}

\newcommand {\be} {\begin{equation}}
\newcommand {\ee} {\end{equation}}
\newcommand {\mult} [1] {\multicolumn{1}{c}{#1}}
\newcolumntype{d}[1]{D{.}{.}{#1}}

\title[Jet production efficiency in FR\,II RGs and QSOs]{On the efficiency of jet production in FR\,II radio galaxies and quasars}

\author[K. Rusinek et al.]{Katarzyna Rusinek$^{1,2}$\thanks{E-mail: krusinek@camk.edu.pl (KR)},
Marek Sikora$^{2}$\thanks{E-mail: sikora@camk.edu.pl (MS)},
Dorota Kozie{\l}-Wierzbowska$^{3}$,
\newauthor Leith Godfrey$^{4}$
\\
$^{1}$Toru\'n Centre for Astronomy, Faculty of Physics, Astronomy and Informatics, Nicolaus Copernicus University,\\~~Grudzi\k{a}dzka 5, 87-100 Toru\'n, Poland\\
$^{2}$Nicolaus Copernicus Astronomical Center, Polish Academy of Sciences, Bartycka 18, 00-716 Warsaw, Poland\\
$^{3}$Astronomical Observatory, Jagiellonian University, ul. Orla, 30-244 Krak\'ow, Poland\\
$^{4}$ASTRON, the Netherlands Institute for Radio Astronomy, Postbus 2, 7990 AA, Dwingeloo, The Netherlands
}

\date{21 December 2016}

\pubyear{2016}

\begin{document}
\label{firstpage}
\pagerange{\pageref{firstpage}--\pageref{lastpage}}
\maketitle

\begin{abstract}
Jet powers in many radio galaxies with extended radio structures appear to exceed their associated accretion luminosities. In systems with very low accretion rates, this is likely due to the very low accretion luminosities resulting from radiatively inefficient accretion flows. In systems with high accretion rates, the accretion flows are expected to be radiatively efficient, and the production of such powerful jets may require an~accretion scenario which involves magnetically arrested discs (MADs). However, numerical simulations of the MAD scenario indicate that jet production efficiency is large only for geometrically thick accretion flows and scales roughly with $(H/R)^2$, where $H$ is the disc height and $R$ is the distance from the BH. Using samples of FR\,II radio galaxies and quasars accreting at moderate accretion rates we show that their jets are much more powerful than predicted by the MAD scenario.
We discuss possible origins of this discrepancy, suggesting that it can be related to approximations adopted in MHD simulations to treat optically thick accretion flow within the MAD-zone, or~may indicate that accretion disks are geometrically thicker than the standard theory predicts.

\end{abstract}


%%%%%%%%%%%%%%%%%%%%%%%%%%%%%%%%%%%%%%%%%%%%%%%%%%
%%%%%%%%%%%%%%%%% BODY OF PAPER %%%%%%%%%%%%%%%%%%

\section{Introduction}
\label{sec:introduction}

The radio-loudness of a quasar is defined as the ratio of radio luminosity (typically at 5~GHz) to optical luminosity (typically in the B-band). The radio luminosity of a quasar is related to jet power $P_j$, while the optical luminosity is related to accretion power $\dot M c^2$, where $\dot M$ is the accretion rate. For this reason, the radio-loudness is a proxy for the jet production efficiency defined to be $\eta_j \equiv P_j/(\dot M c^2)$.

The first quasars were discovered following the identification of bright radio sources with point-like optical sources. However, not all quasars have such strong radio emission: in fact, the majority of quasars have been found to be radio-quiet \citep{1989AJ.....98.1195K}. Present-day radio telescopes are able to detect the faint radio emission of radio-quiet quasars \citep[e.g.][and refs. therein]{2015MNRAS.448.2665W}, however, their radio loudness is up to 3-4 orders of magnitude lower than that of the radio loudest AGNs \citep[e.g.][]{2007ApJ...654...99W}. This indicates a large diversity of jet production efficiency. 

There have been several scenarios proposed to explain such a diversity of jet production efficiency. The two  most popular scenarios are the so-called ``spin paradigm'' \citep{1995ApJ...438...62W,2007ApJ...658..815S,2010MNRAS.406..975G,2011MNRAS.410...53F} and the intermittency of jet production \citep{2003ApJ...593..184L,2006MNRAS.372.1366K}. According to the spin paradigm the jets are powered by rotating BHs and the jet production efficiency, $\eta_j$, is assumed to depend predominantly on the value of the BH spin. The drawback of this assumption is that it implies much lower values of BH spin in radio-quiet AGN than indicated by using ``So{\l}tan argument'' \citep{1982MNRAS.200..115S,2002ApJ...565L..75E,2015ApJ...802..102L} and predicted by numerical simulations of cosmological evolution of supermassive BHs \citep{2013ApJ...775...94V}. The intermittent jet production scenario involves transitions between two accretion modes: one associated with a standard viscous accretion discs and another associated with accretion being driven by MHD winds. While this scenario may be attractive to explain intermittent jet activity observed directly in GRS 1915+105 \citep{2003ApJ...593..184L} and the overabundance of compact radio galaxies in flux limited samples \citep{1997ApJ...487L.135R}, such accretion mode transitions are rather difficult to reconcile with the existence of $10^7 - 10^8$ years old jets observed in FR\,II radio sources \citep{1999AJ....117..677B,2008ApJ...676..147B,2009A&A...494..471O,2012ApJ...756..116A} and also with the lack of evidence for remnant radio lobes around radio-quiet quasars (Godfrey et al., in prep). Furthermore, the ``transition'' models predict bimodal distribution of radio-loudness \citep[e.g.][]{NBB05} and this is observed only if ignoring other than FR\,II sources \citep{2007AJ....133.1615L,2011AJ....141...85R}.

Jet production theories are challenged not only by the large spread of radio-loudness, but also by the fact that the jet powers in many radio galaxies reach values comparable to the accretion powers \citep{1991Natur.349..138R,2007MNRAS.374L..10P,2011MNRAS.411.1909F,SikSta13}. 
In order to produce jets with such high efficiency in the Blandford-Znajek mechanism \citep{1977MNRAS.179..433B}, BHs are required not only to be spinning very fast but also to be threaded by a very large magnetic flux. The required level of magnetic flux threading the black hole can only be maintained if it is confined by the ram pressure of the accretion flow. The latter condition implies a magnetically arrested disc (MAD) scenario, in which the innermost portion of the accretion flow is dynamically dominated by the poloidal magnetic field and accretion proceeds via interchange instabilities \citep{2003PASJ...55L..69N,2008ApJ...677..317I,2009ApJ...704.1065P,2011MNRAS.418L..79T,2012MNRAS.423.3083M}. 

Recent studies of the jet powers in a sample of radio selected FR\,II quasars by \cite{vVel13} \citep[see also][]{vVel15} show that the median jet production efficiency in these objects is tens times lower than maximal predicted by the MAD scenario. Such low jet production efficiency in the MAD scenario would require very low median BH spin and this led the authors to conclude that jet production in these systems does not involve magnetically arrested discs. However, the MAD models predict that the jet production efficiency depends not only on the BH spin, but also has a very strong dependence on the geometrical thickness of the accretion flow. According to \cite{2016MNRAS.462..636A} the jet production efficiency at moderate accretion rates, where standard theory predicts very thin accretion discs, should be hundreds of times lower than that obtained from geometrically thick accretion discs. Therefore, due to the strong dependence of jet production efficiency on disc thickness, the problem is actually the opposite of the one claimed by \citeauthor{vVel13}, and can be expressed by the following question: how is it possible to obtain such high jet production efficiency in these radio-loud AGN, despite their apparently moderate accretion rates, and therefore geometrically thin accretion discs. 

In the current work, we demonstrate the presence of high-$\eta_j$ objects at moderate accretion rates, by considering the dependence of $P_j/L_d$ on the Eddington ratio $L_d/L_{\rm Edd}$ (where $L_d$ is the accretion luminosity and $L_{\rm Edd}$ is the Eddington luminosity) for the following radio-loud AGN samples: $z < 0.4$ FR\,II NLRGs \citep{SikSta13} in \S\ref{subsec:FRII_NLRGs_z04}; FR\,II quasars \citep{vVel13} in \S\ref{subsec:FRII_QSO}; $0.9 < z < 1.1$ NLRGs \citep{2011MNRAS.411.1909F} in \S\ref{subsec:NLRGs_09z11}; and a sample of BLRG+RLQ (Broad-Line Radio Galaxies plus Radio-Loud Quasars) compiled by \cite{2007ApJ...658..815S} in \S\ref{subsec:BLRG_RLQ}. The theoretical implications of the presented distributions -- with particular reference to the applicability of the MAD scenario -- are discussed in \S\ref{sec:discussion} and summarized in \S\ref{sec:summary}.  

We have adopted the $\Lambda$ cold dark matter cosmology, with $H_0=70 \rm{km\,s}^{-1}$, $\Omega_m=0.3$ and $\Omega_{\Lambda}$=0.7.

%-------------------------------------------------
\section{Jet production efficiency}
\label{sec:jet_production}

\subsection{Overview}

In order to adequately assess the distribution of radio galaxies and quasars in the $P_j/L_d$ -- $L_d/L_{\rm Edd}$ plane, we have combined four different samples of radio galaxies and quasars. In the following, we describe each of these samples, and the methods used to estimate $P_j$, $L_d$, and $L_{\rm Edd}$ from the available radio and optical data.

\subsection{FR\,II NLRGs at z$\mathbf{<0.4}$}  
\label{subsec:FRII_NLRGs_z04}

This sample contains 207 FR\,II narrow-line radio galaxies extracted from the sample of $z<0.4$ radio galaxies with extended radio structure selected by \cite{SikSta13}. The objects are taken from Cambridge catalogs and matched with the SDSS, FIRST and NVSS catalogs. 
%The redshift limit insures the H$\upalpha$ line to be within the SDSS spectral range. All objects have radio luminosities  at $\nu=1.4$GHz, as taken from FIRST/NVSS.
The sample is presented in Table~\ref{tab:1-FRII_NLRG_z04} (Appendix~\ref{sect:appendix}, as subsequent tables), where additionally to data presented in \citeauthor{SikSta13} we list values of: disc luminosities -- $L_d$; jet powers -- $P_j$; their ratio -- $P_j/L_d$; and the Eddington ratio -- $\lambda_{\rm Edd} \equiv L_d/L_{\rm Edd}$. The disc luminosities $L_d$ are calculated using the H$\upalpha$ emission line luminosity, $L_{\rm H\upalpha}$, which is available for 152 sources, adopting the conversion formula
\be  L_d[{\rm erg\,s^{-1}}] = 7.8 \times 10^{36} L_{\rm H\upalpha}[\rm L_{\odot}] \,  \ee 
\citep{Net09}, which gives the disc luminosity with uncertainty 0.3 dex. The jet powers are calculated using the 1.4\,GHz monochromatic radio luminosity, $L_{1.4}$, along with the scaling relation of \cite{Wil99}
\be  P_{j}[{\rm erg\,s^{-1}}] = 5.0 \times 10^{22} (f/10)^{3/2} (L_{1.4}[{\rm W\,Hz^{-1}}])^{6/7} \, , \label{eqn:willott}  \ee
where we have assumed the radio spectral index between $151$\,MHz and $1.4$\,GHz is $\alpha_r = 0.8$ (using the convention $F_{\nu} \propto \nu^{-\alpha}$). The formula is based on calorimetry of radio lobes and $f$ is the parameter accounting for errors in the model assumptions. According to \citet{2000AJ....119.1111B} the value of $f$ is between 10 and 20. More secure determination of jet power in FR\,II radio sources is based on the model of hotspots \citep{2013ApJ...767...12G}. Unfortunately hotspots are often very weak or not visible. However, comparing  jet powers of luminous FR\,II sources obtained using the hotspots and radio lobe calorimetry allowed us to calibrate the \citeauthor{Wil99} formula. For luminous FR\,II sources this gives $f\simeq 10$ and uncertainty of $P_j$ calculated using Equation~\ref{eqn:willott} is about 0.3 dex.
%We note that due to the uncertainty in the relation between jet power and radio luminosity, there may be some degree of systematic error introduced by the application of the above relation: \citet{godfrey16} present a detailed analysis of the various empirically and theoretically derived jet power -- radio luminosity scaling relations, and describe in detail the uncertainty surrounding the application of the Willott relation. 
 
\begin{figure}
	\centering
    \includegraphics[width=1.0\columnwidth]{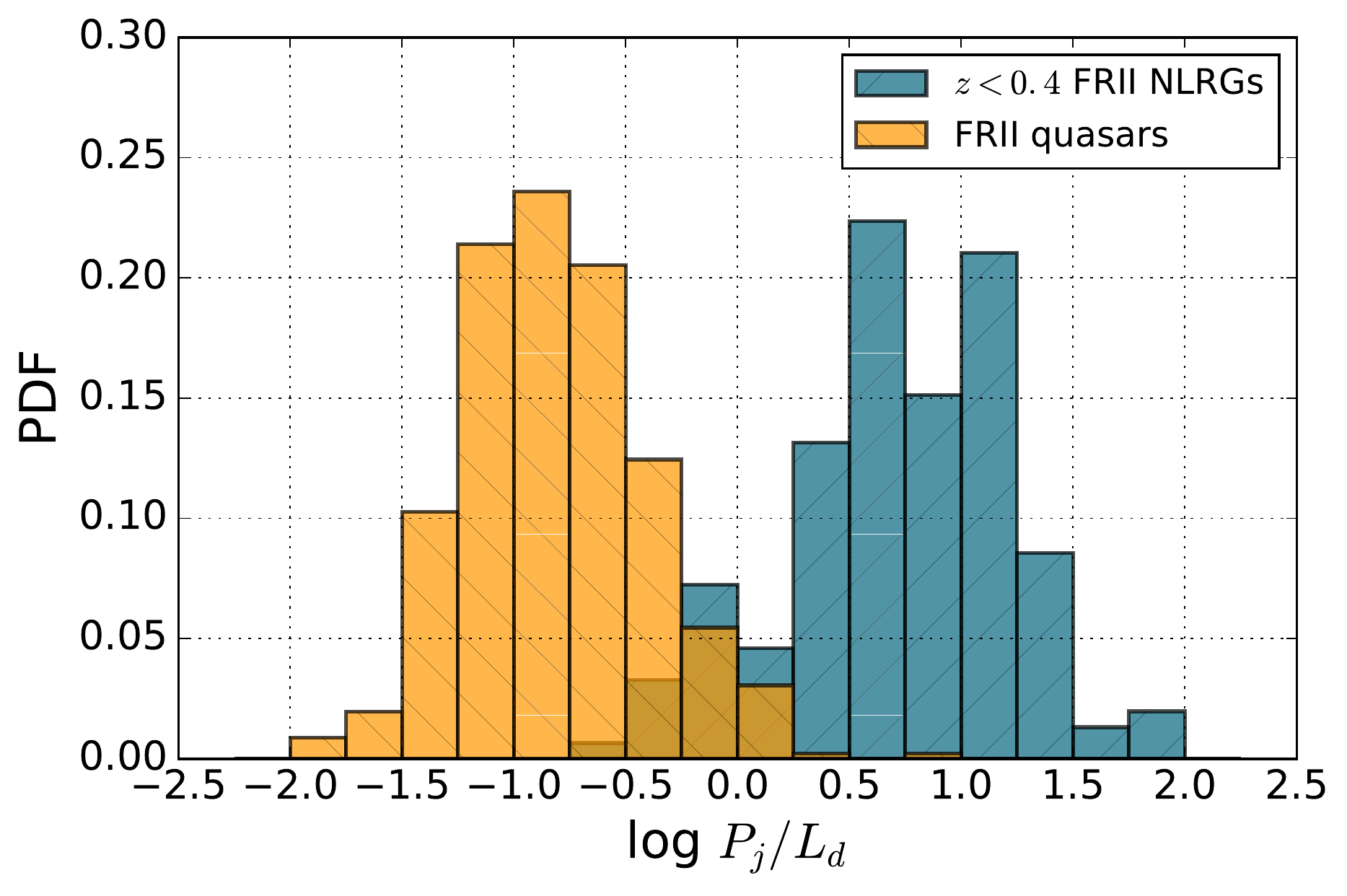}
    \caption{The distribution of the jet efficiency $P_{j}/L_{d}$ for $z<0.4$ FR\,II NLRGs (152 sources) and FR\,II quasars (458 sources) samples represented by blue and orange color respectively. The histogram has been normalised so that the sum of the bin heights is equal to unity.}
    \label{fig:norm_hist_VS} 
\end{figure}

The distribution of $P_j/L_d$ for this sample is plotted in Fig.~\ref{fig:norm_hist_VS}. For many objects $P_j/L_d > 10$, which for disc radiation efficiency $\epsilon_d \equiv L_d/(\dot M c^2) = 0.1$ implies jet production efficiency $\eta_j \equiv \epsilon_d (P_j/L_d) > 1$, where $\dot M$ is the accretion rate. 

%MEDIAN value of P_j/L_d for $\lambda_{\rm Edd} > 0.003$ is:  2.6503545 (67 sources)
%MEDIAN value of P_j/L_d for all sample is:  5.57301015 (152 sources)
 
\begin{figure}
	\centering
    \includegraphics[width=1.0\columnwidth]{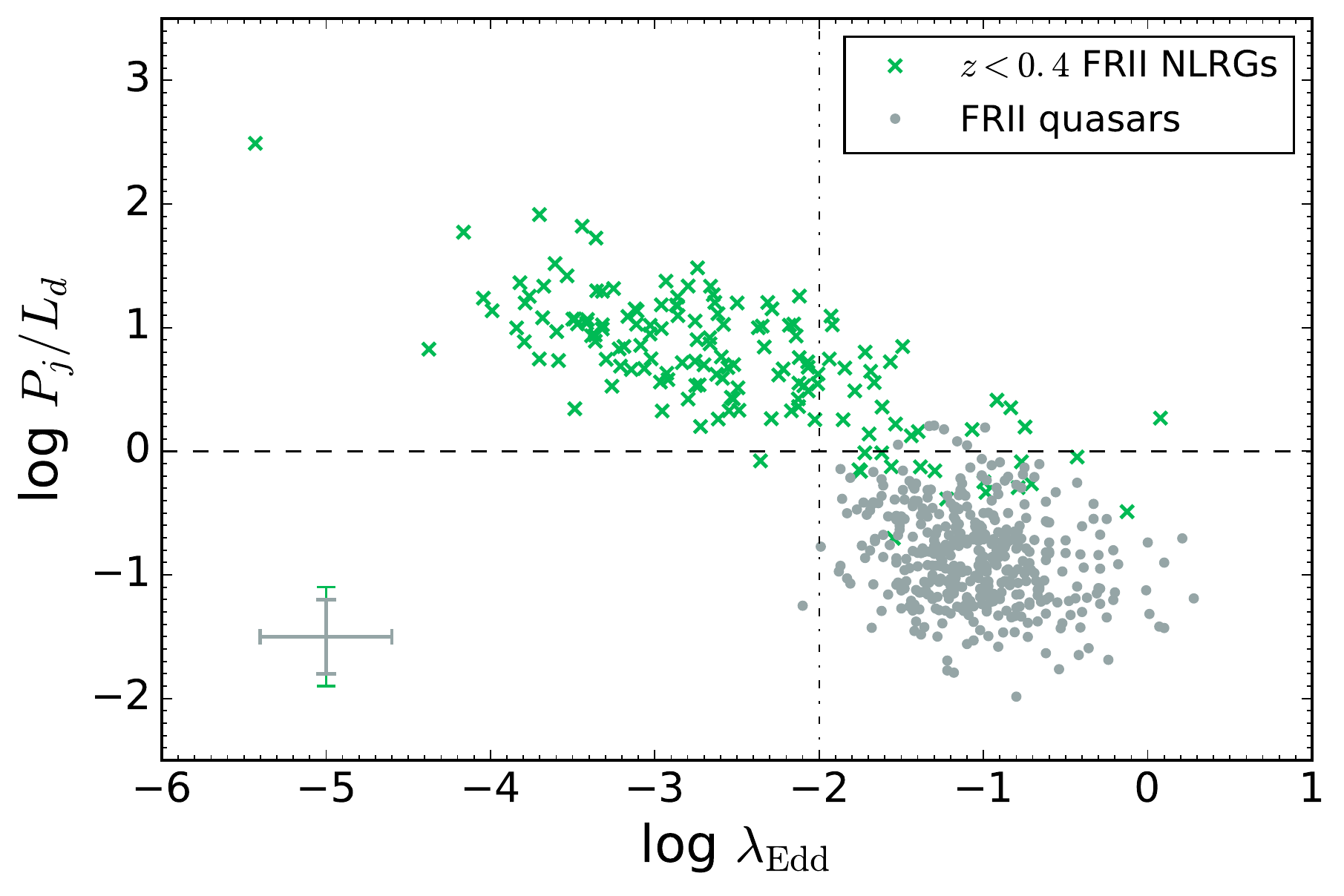}
    \caption{The dependence of $P_{j}/L_{d}$ ratio on the Eddington ratio~$\lambda_{\rm Edd}$. The $z<0.4$ FR\,II NLRGs sample is shown by green crosses while grey dots are for FR\,II quasars sample (here 414 sources). Uncertainties of $P_j/L_d$ and $\lambda_{\rm Edd}$ (described in the respective subsections) are presented in the lower left corner. The horizontal dashed line corresponds to the level where $P_j$ equals $L_d$ and the vertical dot-dashed line marks an approximate value of the Eddington ratio at which the accretion mode is changing from the radiatively inefficient (left side) to the radiatively efficient (right side) \citep{2012MNRAS.421.1569B,2014MNRAS.440..269M}. An apparent anti-correlation exists between these two plotted properties.}
    \label{fig:logEdd_logeta_VS} 
\end{figure}

\begin{figure}
	\centering
    \includegraphics[width=1.0\columnwidth]{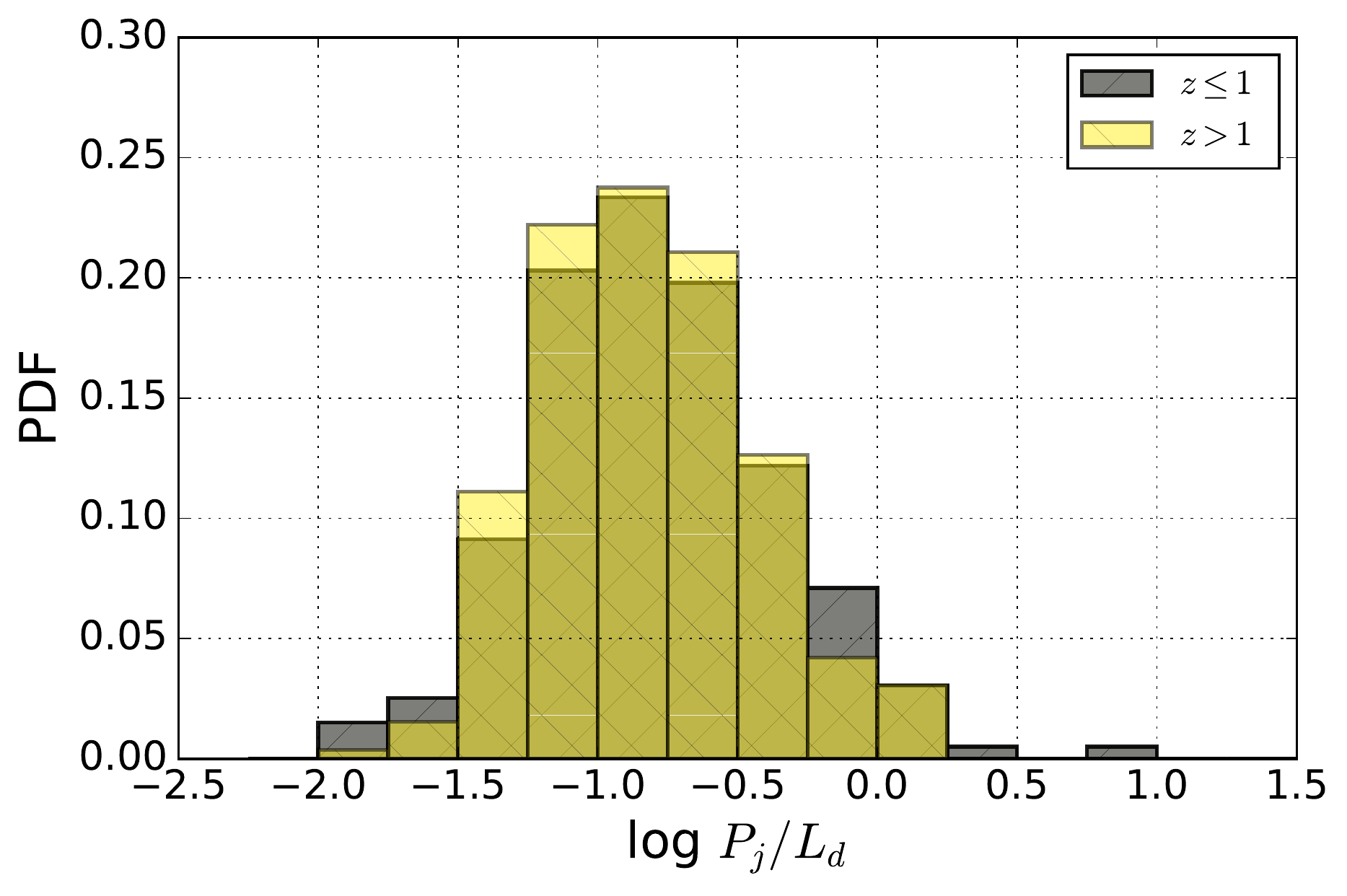}
    \caption{The division of FR\,II quasars sample (458 sources as it is in Fig.~\ref{fig:norm_hist_VS}) based on the boundary value of redshift $z=1$. No~discrepancy between sources with lower (197 objects marked by grey color) and higher (261 objects represented by yellow color) values is present.}
    \label{fig:hist_V_z} 
\end{figure}

In Fig.~\ref{fig:logEdd_logeta_VS} we plot the ratio $P_j/L_d$ against the Eddington scaled accretion luminosity, or Eddington ratio, $\lambda_{\rm Edd} \equiv L_d/L_{\rm Edd}$. As can be seen in this figure, the extreme efficiencies with $P_j/L_d > 10$ and hence $\eta_j > 1$ are possessed only by radio galaxies with very low Eddington ratios, which are therefore presumably operating in the radiatively inefficient accretion regime. The median value of $P_j/L_d$ at moderate accretion rates corresponding to $\lambda_{\rm Edd} > 0.003$ is $2.65$, implying a~modest jet production efficiency of order $\eta_j \sim 0.265 \, (\epsilon_d/0.1)$. Marked in the lower left corner of Fig.~\ref{fig:logEdd_logeta_VS} are the uncertainties for $P_j/L_d$ and $\lambda_{\rm Edd}$. These are calculated based on the uncertainties for $P_j$, $L_d$ and $M_{\rm BH}$ and noting that standard deviations of ratios (and products) of two independently determined quantities, $\sigma_{X/Y} = \sqrt{\sigma_X^2 + \sigma_Y^2}$. The uncertainties of $P_j$, $L_D$ and $M_{\rm BH}$ are estimated to be approximately 0.3 dex \citep[for the latter see][]{2002ApJ...574..740T}, resulting in 0.4 dex uncertainties of $Pj/L_d$ and of $\lambda_{\rm Edd} \propto L_d/M_{\rm BH}$.

\subsection{The FR\,II quasar sample}
\label{subsec:FRII_QSO}

The FR\,II quasar sample used in this work was first obtained by \citet{vVel15} based on the selection of double-lobed radio sources from the FIRST survey catalog, and cross-matching with SDSS quasars.  
%authors established series of limitations including maximum angular separation, flux limit or background substraction. Thereafter, this set was carefully matched to \textit{Sloan Digital Sky Survey} (SDSS) data. More details about the selection algorithm are described in their paper. Our dataset was composed using their Table A1 gathered the complete sample. Firstly, we chose sources targeted based on optical properties what gave us 735 objects. Second step was to eliminate all objects which, according to V15 criterions, did not pass quality cut. Finally, set from V15 contains 458 sources.
In Table~\ref{tab:2-FRII_QSO} we present the relevant data for this sample, including the monochromatic rest-frame luminosity at 1.4\,GHz, $L_{1.4}$, and~if available, masses of black holes and Eddington ratios. The radio luminosities were calculated based on the 1.4\,GHz lobe flux densities given by \citeauthor{vVel15}, and k-corrected assuming radio spectral index $\alpha_r=0.85$, along with standard $\Lambda$CDM cosmology, as specified in Section \ref{sec:introduction}. The jet power $P_j$ was calculated using Equation \ref{eqn:willott}. The black hole masses and Eddington ratios, when available, were taken from \citet{She11} thereby reducing the sample from 458 to 414 sources. 

%In order to determine lacking black hole masses in V15 sample, we used an online available catalog of quasar properties from SDSS Data Release 7 (DR7) made by \cite{She11}. It involves an ample compilation of properties of 105 783 quasars including continuum and emission line measurements, radio properties, flags indicating broad absorption line quasars, disc emitters and virial black hole mass estimated based on various calibrations. Specfic description of construction of this catalog and discussion of its limitations may be found in the paper. To compare our V-sample with \cite{She11} set we extracted SDSS coordinates. Finally, after rejection all sources not marked as lobe-dominant objects (radio flag from FIRST), 414 sources were received. 

The $P_j/L_d$ histogram and dependence of $P_j/L_d$ on $\lambda_{\rm Edd}$ for this sample of FR\,II quasars are plotted together with $z<0.4$ FR\,II NLRGs in Fig.~\ref{fig:norm_hist_VS} and Fig.~\ref{fig:logEdd_logeta_VS}. As can be seen, the median jet production efficiency in FR\,II quasars is $\sim 0.02 (\epsilon_d/0.1)$, i.e.~$\sim 13$ times lower than in the $\lambda_{\rm Edd} > 0.003$ subsample of $z<0.4$ FR\,II NLRGs. 

%MEDIAN value of P_j/L_d for all sample is:  0.1480688  (458 sources)
%MEDIAN value of P_j/L_d for M_BH sample is: 0.1374292  (414 sources)

In Fig.~\ref{fig:hist_V_z} we show the $P_j/L_d$ distributions for the FR\,II quasars divided into two subsamples, with $z>1$ and $z<1$. The fact that the $P_j/L_d$ distributions are very similar for the high- and low-redshift subsamples indicates that the difference in median $\eta_j$ between FR\,II quasars and $z<0.4$ NLRGs at $\lambda_{\rm Edd} > 0.01$ is not caused by cosmological evolution of jet production efficiency, but rather by the different flux limits and procedures to select the two samples. 

Uncertainties of $L_d$ luminosities derived in \citet{She11} using bolometric corrections to optical luminosity at $5100$\,\AA~are about 0.1 dex \citep{2006ApJS..166..470R}, while uncertainties of $M_{\rm BH}$ derived by \citeauthor{She11} using virial estimators are $\sim0.4$ dex. With these uncertainties and 0.3 dex uncertainty of $P_j$, the uncertainties of $P_j/L_d$ and $\lambda_{\rm Edd}$ for FR\,II quasars are $\sim0.3$ dex and $\sim0.4$ dex, respectively. They are marked, together with uncertainties for the $z < 0.4$ FR\,II NLRGs sample, in the lower left corner of Fig.~\ref{fig:logEdd_logeta_VS}.

%This led van Velzen \& Falcke (2013) to conclusion, that  MAD scenario cannot operate in FRII quasars, because would require too small BH spins. However, before addressing this issue, we will verify how much presented differences can be associated with selection effects. It can be done by adding to our plots some other, albeit very limited, samples which are based on different selection criteria.

\subsection{$\mathbf{0.9 < z < 1.1}$ sample of NLRGs}
\label{subsec:NLRGs_09z11}

As we can see from Fig.~\ref{fig:logEdd_logeta_VS}, the sample of $z<0.4$ FR\,II NLRGs is poorly represented at $\log \lambda_{\rm Edd} > -1.5$. This is primarily due to a low number of high accretion rate AGNs with very massive BHs in the local Universe. In order to verify how much this incompletness affects the average jet production efficiency in FR\,II RGs, we extend the FR\,II NLRGs sample by adding 27 $z \sim 1$ NLRGs selected in $0.9 < z < 1.1$ taken from \citet{2010MNRAS.405..347F}. With a few exceptions, they all have FR\,II radio morphologies. The relevant data for this sample are presented in the Table~\ref{tab:3-NLRG_09z11}, which are taken from Table~1 in \citet{2011MNRAS.411.1909F}, as well as Table~3 in \citet{2015MNRAS.447.1184F}. As before, $P_j$ is calculated based on Equation \ref{eqn:willott}. The disc luminosity for this sample has been calculated using mid-IR data from {\it Spitzer Space Telescope} at wavelength $12\,{\rm \upmu m}$ along with the following scaling relation: $L_{d} = 8.5 \times [\nu L_{\nu}]$ \citep{2006ApJS..166..470R}. Black hole masses have been derived using relation between the black hole and bulge masses \citep{2004ApJ...604L..89H}. Their uncertainty is $\sim0.3$ dex. Combining it with uncertainty of disc luminosities, $\sim0.2$ dex, and jet powers, $0.3$ dex, gives uncertainties of $P_j/L_d$ and of $\lambda_{\rm Edd}$, both about $0.4$ dex.

\begin{figure}
	\centering
    \includegraphics[width=1.0\columnwidth]{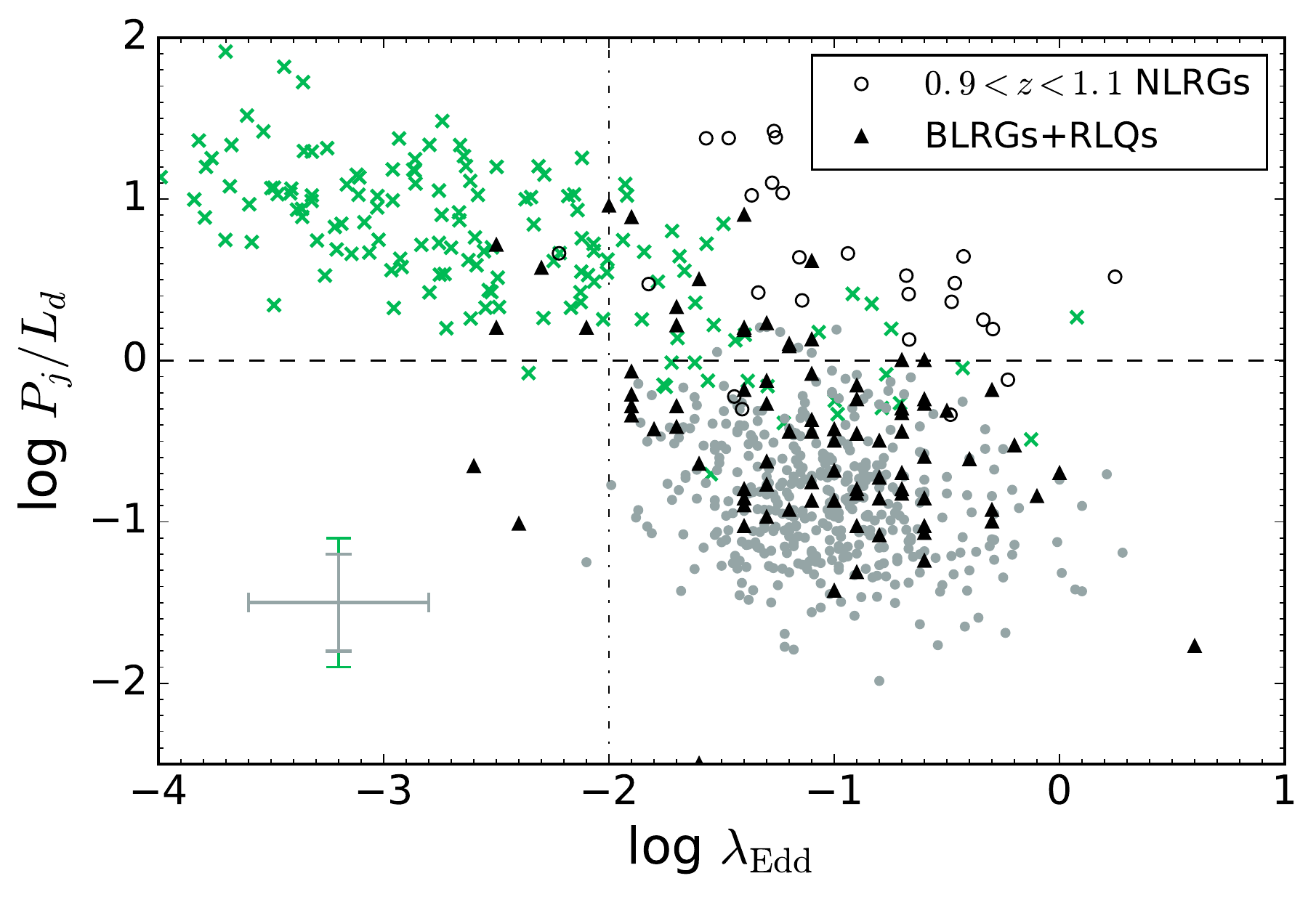}
    \caption{The same plot as in the Fig.~\ref{fig:logEdd_logeta_VS} with two added samples: $0.9 < z < 1.1$ NLRGs as empty circles (27 sources); BLRG+RLQ as black triangles (87 sources). Uncertainties of $P_j/L_d$ and $\lambda_{\rm Edd}$ for all four samples are presented in the lower left corner (their values for $z < 0.4$ FR\,II NLRGs and $0.9 < z < 1.1$ NLRGs and for FR\,II quasars and BLRG+RLQ are equal).}
    \label{fig:logEdd_logeta_4in1} 
\end{figure}

The sample is plotted in Fig.~\ref{fig:logEdd_logeta_4in1}. The median $P_j/L_d$ is similar to that of the $z<0.4$ FR\,II NLRGs sample for $\lambda_{\rm Edd} > 0.003$ and uncertainties of the $z \sim 1$ NLRGs sample are the same. They are marked in the lower left corner.

%MEDIAN value of P_j/L_d is:  3.3024606 (27 sources) 

\subsection{A BLRG+RLQ sample}
\label{subsec:BLRG_RLQ}

The incompletness of SDSS quasars at moderate accretion rates \citep{2013ApJ...764...45K} may introduce a bias in our analysis of FR\,II quasars due to underrepresentation of such objects, particularly at $\lambda_{\rm Edd} < 0.03$. In order to verify whether the incompleteness of SDSS quasars at moderate accretion rates can significantly affect the average value of $\eta_j$ of our sample, we complete our studies of jet production efficiency by adding a sample of broad-line RGs co-selected with low redshift radio-loud quasars. The sample is comprised of radio-loud broad-line AGN with redshift $z<0.4$, selected from \citet{1989ESOSR...7....1V} by \citet{1994ApJS...90....1E,2003ApJ...599..886E} and used by \citet{2007ApJ...658..815S} to study radio-loudness of these objects. Using a formal, luminosity related definition of quasars, these objects were divided by \citeauthor{2007ApJ...658..815S} into two subsamples: broad-line radio galaxies (BLRGs) and radio-loud quasars (RLQs). 
%However, because these subsamples significantly overlap regarding the Eddington ratio values, they are 're-joined' in this paper. 
The~BLRG+RLQ sample data are listed in Table~\ref{tab:4a-BLRG} and \ref{tab:4b-4RLQ}. 
%The Table is organized in the same way as Tables 1 and 2 in Sikora et al. (2007), but with the addition of two columns, $\log P_j$ and $\log P_j/L_d$. 
As with the previous samples, $P_j$ is calculated using Equation~\ref{eqn:willott}, but in this case, we have had to extrapolate flux densities at 5\,GHz to 1.4\,GHz using a radio spectral index $\alpha_r = 0.8$. The disc luminosity is calculated based on the B-band and using the bolometric correction from \citet{2006ApJS..166..470R}. Its uncertainty is $\sim 0.1$ dex. Black hole masses have been derived using virial estimators \citep[e.g.][]{2002ApJ...579..530W}, and uncertainties of such estimators are $\sim 0.4$ dex. With these uncertainties and $0.3$ dex uncertainty of $P_j$, the uncertainties of $P_j/L_d$ and $\lambda_{\rm Edd}$ are $\sim 0.3$ dex and $\sim 0.4$ dex, respectively. They are marked in Fig.~\ref{fig:logEdd_logeta_4in1} in the lower left corner together with uncertainties for the $z \sim 1$ NLRGs sample. As we can see in Fig.~\ref{fig:logEdd_logeta_4in1}, despite these large uncertainties, the BLRG+RLQ sample is fully consistent with the sample of FR\,II quasars from \citeauthor{vVel15}
%however, as discussed in section~\ref{subsec:FRII_QSO}, the  NLRGs  samples tend to have higher %median $P_j/L_d$ due to the fact that they are selected using radio catalogs which have, on %average, much higher flux limits than the FIRST catalogue used by \citeauthor{vVel15} to select %FR\,II quasars. 
%MEDIAN value of P_j/L_d for all sample is:  0.3541133 (87 sources) 
%MEDIAN value of P_j/L_d for BLRG is:  0.706549 (37 sources) 
%MEDIAN value of P_j/L_d for QSOs is:  0.1746056 (50 sources) 

%\vspace{2.0cm}
%-------------------------------------------------
\section{Discussion}
\label{sec:discussion}

The applicability of the MAD scenario for the production of powerful AGN jets was recently investigated using 3D general-relativistic, magnetohydrodynamic simulations \citep{2008ApJ...677..317I,2009ApJ...704.1065P,2011MNRAS.418L..79T,2012MNRAS.423.3083M}. These studies demonstrated that magnetically arrested discs have the ability to launch jets with a power comparable to the accretion power, as is required to explain the energetics of radio lobes in the radio-loudest FR\,II sources \citep{1991Natur.349..138R,2007MNRAS.374L..10P,2011MNRAS.411.1909F,SikSta13}.

However, the MAD simulations indicate there is a clear trend of decreasing efficiency of relativistic jet production with decreasing geometrical thickness of the accretion flow. This trend was found in the case of ``non-radiative'' models with disk thickness $H/R = 1.0 - 0.3$ \citep{2011MNRAS.418L..79T,2012MNRAS.423.3083M}, and was explained as the being due to loading boundary layers of the Blandford-Znajek outflow with mass to such a level that these outflows do not gain relativistic speeds. More recently, simulations of the MAD scenario were extended by \citet{2016MNRAS.462..636A} to cover the case of the optically thick accretion flows, with $H/R \sim 0.1$. Combining the results of these simulations with results obtained for non-radiative and geometrically thicker accretion flows they derived the following empirical formula describing a dependence of the jet production efficiency $\eta_j$ on geometrical thickness and dimensionless BH spin $a$,
\be \eta_j \simeq 4 a^2 \, \left( 1+ \frac{0.3 a}{1 + 2 (H/R)^4} \right)^2
\, (H/R)^2 \,, \ee
which for  $H/R \ll 1$ gives $\eta_j \sim 4 a^2 (1+0.3 a)^2 (H/R)^2$. According to the standard accretion disc model \citep{1973blho.conf..343N,1989MNRAS.238..897L}, maximal thickness of a disc accreting onto 
a BH with $a \sim 1$ and producing radiation at a rate $\lambda_{\rm Edd} \sim 0.1$ 
is $H/R \sim 0.04$. For these parameters the above formula gives $\eta_j \simeq 0.01$. This is a factor 2 less than the median value of the FR\,II quasars
sample and by a factor 20 less than its upper bound in the $P_j/L_d$ vs.~$\lambda_{\rm Edd}$
plots. Noting $\sim0.4$ dex uncertainties of $P_j/L_d$, it is 
rather unlikely that above discrepancy is resulting from errors 
of $P_j$ and/or $L_d$. Then we can envisage two possible solutions of this 
discrepancy. One is that because MHD simulations of radiative, optically thick 
accretion flows are still not fully self-consistent, the extrapolation
of dependence of $\eta_j$ on $H/R$ indicated by non-radiative accretion 
flows down to the regime of optically thick accretion flows can be quantitatively
inaccurate. Another possibility is that optically thick accretion discs are much 
thicker than predicted by the standard accretion disc theory. The disc can be 
thicker in presence of strong toroidal magnetic fields \citep[e.g.][]{2007MNRAS.375.1070B,2016MNRAS.462..960S}, or can be accompanied by heavy, viscously driven corona \citep{2015A&A...580A..77R,2015ApJ...809..118B}. Furthermore, within the MAD zone the disc is radially balanced against gravity by dynamically dominated poloidal magnetic fields and, therefore, even if outside the MAD-zone the disc is geometrically thin, within the MAD-zone it can become sub-Keplerian and thicker than the standard one. The suspicion that approximations used by \citeauthor{2016MNRAS.462..636A} to treat in MHD simulations the optically thick disc are inaccurate is also supported by the fact, that they predict larger radiative efficiency of MADs than of standard accretion discs, whilst 
observations indicate the opposite: radio-quiet quasars have been found to be more luminous in UV than radio-loud quasars \citep[][and refs.~therein]{2016MNRAS.459.4233P}.

Obviously, not all radio-loud quasars have FR\,II radio morphology. According to \citet{2006AJ....131..666D}, most of them have radio structures too compact to be resolved, or if resolved, are recorded as CSOs (Compact Symmetric Objects), CSS (Compact Steep-Spectrum) sources and GPS (GHz-peaked spectrum) sources \citep[][and refs. therein]{2012ApJ...760...77A}. Many of them are as radio loud as FR\,II RGs and quasars and therefore can also be considered to involve MAD scenario. However, about $90$ percent of all quasars are not detected in radio or have very weak radio emission which can be
%at a level that is low enough to be produced by non-MAD models. Such jets can still be produced by the %Bladford-Znajek mechanism. Nonetheless, their radio emission can be also 
associated with starburst activities \citep{2011ApJ...739L..29K} or with shocks formed by the quasar driven outflows \citep{2014MNRAS.442..784Z}.

\begin{figure*}
	\centering
    \includegraphics[width=0.68\textwidth]{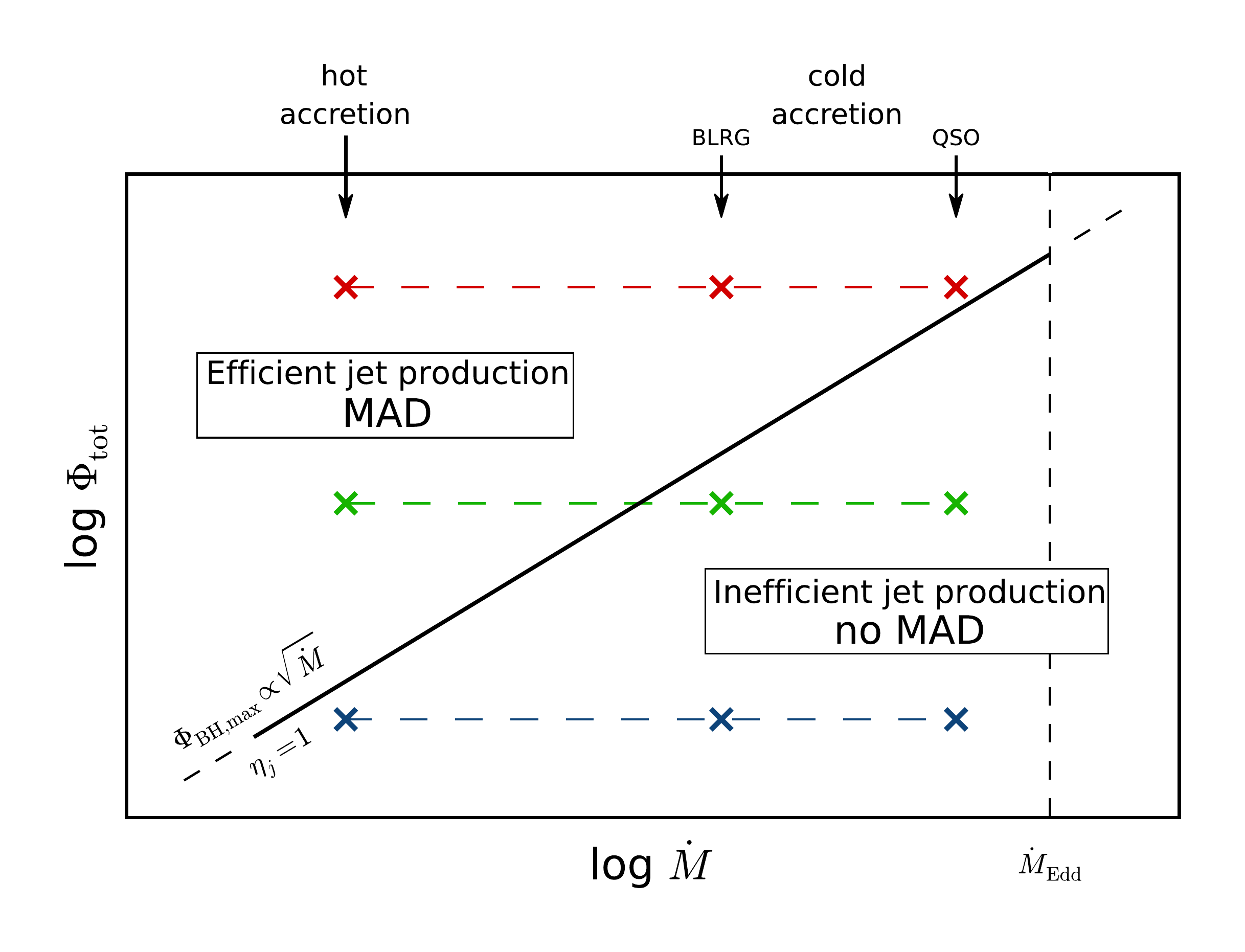}
    \caption{This figure illustrates a condition that must be satisfied in order to obtain a magnetically arrested disc, and also demonstrates how this condition dictates the fraction of radio loud AGN as a function of accretion rate, $\dot{M}$. The inner accretion flow will become magnetically arrested only if $\Phi_{\rm tot}$ exceeds $\Phi_{\rm BH,max}(\dot{M})$, where $\Phi_{\rm tot}$ is the magnetic flux assumed to be accumulated in the central region of an AGN following the hot accretion phase preceding the higher accretion event associated with the BLRG or quasar phenomenon, while $\Phi_{\rm BH,max}(\dot{M})$ is the maximal magnetic flux that can be confined on a BH by an accretion flow. For $\Phi_{\rm tot} < \Phi_{\rm BH, max}(\dot{M})$ the magnetic flux will be entirely enclosed on the black hole and the magnetically arrested disc will not be formed. It is assumed that efficient jet production (and therefore highly radio loud AGN) only occur in the MAD case, when $\Phi_{\rm tot} > \Phi_{\rm BH,max}(\dot{M})$, while jet production is assumed to be inefficient in the case of No MAD when $\Phi_{\rm tot} < \Phi_{\rm BH,max}(\dot{M})$. This condition implies that the fraction of AGN that are radio loud decreases with increasing $\dot{M}$. For details see \citet{2013ApJ...764L..24S}.}
    \label{fig:mag_flux_vs_accretion_rate}  
\end{figure*}

It is tempting to speculate, that the reason for a~very small fraction of radio-loud quasars is associated with ``a~steep magnetic-flux function'' of quasar precursors developed during a hot, quasi-spherical accretion  phase, where the steepness can be determined by different levels of ordering in the magnetic fields that are advected to the center, and/or by the duration of the quasar pre-phase 
\citep{2013ApJ...764L..24S}. This scenario is schematically illustrated in Fig.~\ref{fig:mag_flux_vs_accretion_rate}. For a given amount of magnetic flux $\Phi_{{\rm tot}}$ accumulated during the quasar pre-phase (Bondi accretion phase), the MAD accretion mode will operate during a subsequent BLRG/quasar phase only if $\Phi_{{\rm tot}}$ exceeds $\Phi_{{\rm BH,max}}(\dot M)$. One can easily deduce from this figure, that the fraction of objects operating in the MAD mode is predicted to increase with decreasing accretion rate. Such a trend is indicated at higher accretion rates by studies of quasars \citep{2015AJ....149...61K}, at moderate accretion rates -- by studies of double-peak emission line galaxies \citep{2004ApJ...614...91W} , and at very low accretion rates -- by studies of nearby galaxies \citep{2003ApJ...583..145T,2005ApJ...625..716C}.
Indications of a possible MAD scenario operation during the Bondi accretion phase have been recently provided  by studies of $P_j/\dot M_{{\rm Bondi}}$ in several nearby radio galaxies \citep{2015MNRAS.449..316N}.

Finally, we consider variability of the accretion rate as a possible complicating factor in our interpretation of the $P_{j}/L_{d} - \lambda_{\rm Edd}$ distribution. The radio luminosity is related to the total energy content of the lobes, and is dependent on the time-averaged jet power averaged over the lifetime of the source. As a result, the jet power calculated from the lobe radio luminosity represents a measure of the time-averaged jet power. The hotspot luminosity may vary on short timescales due to variation in jet power, but the hotspots typically contribute only a small fraction of the total radio luminosity \citep{2008MNRAS.390..595M}, and so the integrated lobe luminosity will not be significantly affected by short timescale variation in the jet power. In contrast, the disc luminosity is a~measure of instantaneous accretion rate, and the accretion rate may vary significantly on timescales much shorter than the lifetime of the radio galaxy. As a result, variability of the accretion disc luminosity will cause variability in the ``apparent" jet production efficiency and Eddington ratio. Consider for example a source in which the accretion power varies by a factor of 10 between its maximum and minimum accretion rates. This object, if observed during its accretion rate minimum, will appear to have 10 times lower Eddington ratio and 10 times larger jet production efficiency than if it were observed at its accretion rate maximum. In effect, variability of the accretion rate will cause the $P_{j}/L_{d} - \lambda_{\rm Edd}$ distribution to be stretched along a line with slope $-1$ in the $P_{j}/L_{d} - \lambda_{\rm Edd}$ plane, broadly consistent with the slope of the distribution of points shown in Figure~\ref{fig:logEdd_logeta_4in1}. Furthermore, for a duty cycle $\sim 1/2$, the object's apparent jet production efficiency will be about 5 times larger than the true jet production efficiency when observed at its minimum accretion rate, and about 2 times smaller when observed at its maximum accretion rate. A natural driver of variability in the accretion rate is viscous instabilities in accretion discs \citep{2002ApJ...576..908J,2011MNRAS.414.2186J}. Observational support for this hypothesis may come from the spatial modulation of the radio brightness distributions seen in some large scale jets \citep{2012ApJ...758L..27G}.

%\textcolor{red}{We note that the observed drop in average jet production efficiency for radio-loud AGNs is much smaller than the difference between radio-loudness of X-ray binaries (XRBs) in the low/hard and in high/soft states. This difference between AGN\textcolor{cyan}{s} and XRBs can be explained assuming that large scale magnetic fields involved in the jet launching process are developed by dynamo processes \citep{2004ApJ...613..716C}, rather than being advected from external region as has previously been assumed in the case of radio-loud AGNs \citep[e.g.][]{2013ApJ...764L..24S,2016arXiv161004061C}.  
%AGN\textcolor{cyan}{s} with sub-MAD magnetic fluxes are expected to form a similar pattern in the 'radio-loudness -- Eddington ratio' plane as XRBs. Such sub-MAD AGNs are likely represented by the lower radio-loudness pattern seen in Fig. 3 of \citealp{2007ApJ...658..815S} 
%(see also \citealp{2002ApJ...564..120H}). At very low accretion rates these sub-MAD AGNs, similar to  XRBs in low-hard states, are associated with compact, steady radio sources. At higher accretion rates most of them are radio-quiet, and those which are radio-loud have compact radio structures. Some of them can be FR\,II-family babies, others presumably are short lived sources produced in similar stochastic manner as radio ejecta observed in some XRBs \citep{2014SSRv..183..323F}.}

%-------------------------------------------------
\section{Summary}
\label{sec:summary}

The compilation of data on $P_j/L_d$ and $\lambda_{\rm Edd}$ taken from four independently selected samples clearly show a~drop of the jet production efficiency at higher accretion rates (Figure~\ref{fig:logEdd_logeta_4in1}). It is tempting to connect this drop in jet production efficiency with a transition from radiatively inefficient, optically thin accretion flows (RIAF) to the standard, optically thick accretion discs and assume that the key ingredient for jet production is the presence of the hot gas associated with the bulk accretion flow (at low accretion rates) or with the disc corona (at high accretion rates) \citep[e.g.][]{2004ApJ...613..716C,2013ApJ...770...31W}. However, in order to produce at $\lambda_{\rm Edd} > 0.01$ jets with $P_j/L_d$ approaching unity, as is observed for many objects, requires magnetic fluxes in the central regions too large to be supported by disc coronas. 
This argument favors the MAD-scenario, but with a geometrical thickness of accretion flows much larger than the thickness predicted by standard theory. Whether the discs become thicker once entering the MAD-zone, or must be already thicker prior to the MAD-zone is the subject for future investigations.

%-------------------------------------------------
\section*{Acknowledgements}

We thank an anonymous referee for his/her critical comments which helped us to improve the paper. MS thanks Aleksander S\c adowski for helpful discussions.
The research leading to these results has received funding from the Polish National Science Centre grant 2013/09/B/ST9/00026 and from the European Research Council under the European Union's Seventh Framework Programme (FP/2007-2013) / ERC Advanced Grant RADIOLIFE-320745.


%%%%%%%%%%%%%%%%%%%%%%%%%%%%%%%%%%%%%%%%%%%%%%%%%%
%%%%%%%%%%%%%%%%% APPENDICES %%%%%%%%%%%%%%%%%%%%%

\appendix

\section{Samples}
\label{sect:appendix} 

Here we present astrophysical properties of our samples with detailed calculations described in Section~\ref{sec:jet_production}. Complete tables are available as~a~supplementary material in the online journal. A portion is shown here for guidance regarding its form and content.

% Don't change these lines
\bsp	% typesetting comment
\label{lastpage}

\newgeometry{margin=2cm} % modify this if you need even more space
\begin{landscape}

\begin{table}
	\centering                                                                                                               
	\caption{Radio and optical properties of $z<0.4$ FR\,II NLRGs from Table 1 in \citet{SikSta13} with some calculated values in this work. The disc luminosities $L_d$ were determined using $L_{{\rm H} \upalpha}$.}
	\label{tab:1-FRII_NLRG_z04}
	\begin{tabular}{*{11}{c}}
		\hline
		SDSS ID & Cambridge Cat.~ID & Redshift & $\log L_{1.4}$ & $\log L_{\mathrm{H}\upalpha}$ & $\log L_{\mathrm{[O\,III]}}$ & \mult{$\log L_{d}$} & \mult{$\log P_{j}$} & $\log P_{j}/L_{d}$ & $\log M_{\mathrm{BH}}$ & $\log \lambda_{\rm Edd}$ \\
		& & & [WHz$^{-1}$] & [${\rm L_{\odot}}$] & [${\rm L_{\odot}}$] & \mult{[ergs$^{-1}$]} & \mult{[ergs$^{-1}$]} & & [${\rm M_{\odot}}$] & \\ 
		\hline \hline
		0312.51689.471 & 4C +00.56            & 0.0524 & 25.34 & 7.605  & 7.572 & 44.497 & 44.4190 &-0.0781 & 8.74 & -2.3568  \\
		0349.51699.169 & 6C B165818.4+630042  & 0.1063 & 25.45 & 6.417  & 6.579 & 43.309 & 44.5133 & 1.2042 & 7.83 & -2.6348  \\
		0366.52017.349 & 6C B171944.8+591634  & 0.2212 & 25.59 & 7.486  & 6.889 & 44.378 & 44.6333 & 0.2552 & 8.29 & -2.0258  \\
		0432.51884.345 & 7C B073404.1+402639  & 0.3905 & 25.59 & \ldots & 6.806 & \ldots & 44.6333 & \ldots & 8.66 &  \ldots  \\
		0436.51883.010 & 6C B075738.1+435851  & 0.2554 & 25.66 & 6.899  & 6.740 & 43.791 & 44.6933 & 0.9022 & 8.42 & -2.7428  \\
		0439.51877.637 & 7C B081405.1+450809  & 0.1422 & 25.43 & 5.690  & 6.322 & 42.582 & 44.4961 & 1.9140 & 8.17 & -3.7018  \\
		0448.51900.335 & 6C B084421.9+571115  & 0.1937 & 26.08 & 7.515  & 7.887 & 44.407 & 45.0533 & 0.6462 & 7.98 & -1.6868  \\
		0450.51908.330 & 4C +56.17            & 0.1409 & 26.05 & 7.107  & 6.912 & 43.999 & 45.0275 & 1.0284 & 8.04 & -2.1548  \\
		\hline                                                                                                                      
	\end{tabular}
\end{table}

\begin{table}
	\centering                                                                                          
	\caption{Some properties of FR\,II quasars from Table A1 in \citet{vVel15}. Few columns calculated in this work were added, together with black hole masses and Eddington ratios taken from \citet{She11}.}
	\label{tab:2-FRII_QSO}
	\begin{tabular}{*{2}{r}*{8}{c}}
		\hline
		\mult{SDSS RA} & \mult{SDSS Dec} & Redshift & Lobe flux & $\log L_{1.4}$ & $\log L_{d}$ & $\log P_{j}$  & $\log P_{j}/L_{d}$ & $\log M_{\mathrm{BH}}$ &\mult{$\log \lambda_{\rm Edd}$} \\
		\mult{deg} & \mult{deg} & & Jy & [WHz$^{-1}$] & [ergs$^{-1}$] &  [ergs$^{-1}$] & & [$M_{\odot}$] & \\
		\hline \hline
		2.910161  & -10.749515 & 1.2712 & 0.0963 & 26.9061 & 46.5254 & 45.7613 & -0.7641 & 9.65   & -1.19    \\
		6.808142  &   1.610954 & 0.9010 & 0.1056 & 26.5880 & 45.6681 & 45.4887 & -0.1794 & \ldots &  \ldots  \\
		10.165798 &  15.055892 & 0.8844 & 0.0294 & 26.0133 & 45.9686 & 44.9961 & -0.9725 & \ldots &  \ldots  \\
		11.079023 &  -9.002630 & 0.9672 & 0.0509 & 26.3449 & 46.1815 & 45.2803 & -0.9012 & 7.83   &  0.10    \\
		12.273874 &  -0.514230 & 3.2310 & 0.0196 & 27.1639 & 46.3095 & 45.9823 & -0.3272 & \ldots &  \ldots  \\
		13.785633 & -10.868412 & 1.3810 & 0.0303 & 26.4898 & 45.9742 & 45.4045 & -0.5697 & \ldots &  \ldots  \\
		15.872669 &   0.681930 & 1.4331 & 0.1200 & 27.1259 & 46.6872 & 45.9497 & -0.7375 & 9.47   & -0.94    \\
		19.457974 &  -9.098518 & 0.8284 & 0.1041 & 26.4944 & 46.0661 & 45.4085 & -0.6576 & 9.18   & -1.34    \\
		\hline                                                                                                 
	\end{tabular}
\end{table}

\begin{table}
	\centering                                                                                                           
	\caption{Properties from the Table 1 in \citet{2011MNRAS.411.1909F} and Table 3 in \citet{2015MNRAS.447.1184F} with added $\log P_{j}$ and $\log P_{j}/L_{d}$ values.}
	\label{tab:3-NLRG_09z11}
	\begin{tabular}{*{8}{c}}
		\hline
		Cambridge Cat.~ID & Redshift & $\log L_{\nu151{\rm MHz}}$ & $\log L_{d}$ & \mult{$\log P_{j}$} & $\log P_{j}/L_{d}$ & $\log M_{\mathrm{BH}}$ & $\log \lambda_{\rm Edd}$ \\
		& & [WHz$^{-1}$sr$^{-1}$] & [ergs$^{-1}$] & \mult{[ergs$^{-1}$]} & & [${\rm M_{\odot}}$] & \\ 
		\hline \hline
		3C 280    & 0.997 & 28.29 & 46.7070 & 47.2258 & 0.5188 & 8.346 &  0.2467  \\
		3C 268.1  & 0.974 & 28.21 & 45.6890 & 47.1573 & 1.4683 & 7.476 &  0.0993  \\
		3C 356    & 1.079 & 28.12 & 46.4350 & 47.0801 & 0.6451 & 8.746 & -0.4260  \\
		3C 184    & 0.994 & 28.01 & 45.6080 & 46.9858 & 1.3778 & 8.966 & -1.4685  \\
		3C 175.1  & 0.920 & 27.98 & 45.5780 & 46.9601 & 1.3821 & 8.726 & -1.2596  \\
		3C 22     & 0.937 & 27.96 & 46.8130 & 46.9430 & 0.1300 & 9.366 & -0.6676  \\
		3C 289    & 0.967 & 27.95 & 46.2710 & 46.9344 & 0.6634 & 9.096 & -0.9393  \\
		3C 343    & 0.988 & 27.78 & 46.5940 & 46.7887 & 0.1947 & 8.776 & -0.2958  \\
		\hline                                                                                                                      
	\end{tabular}    
\end{table}
\end{landscape}

\begin{landscape}
\begin{table}
	\renewcommand{\thetable}{A\arabic{table}a}
	\small
	\centering                                                                              
	\caption{Some properties of BLRGs from Table 1 in \citet{2007ApJ...658..815S} with calculated properties in this work.}
	\label{tab:4a-BLRG}
	\begin{tabular}{*{7}{c}*{1}{d{3.5}}*{5}{c}}
		\hline 
		IAU & Name & Redshift & $m_{V}$ & $A_{V}$ & $\kappa_{\star}$ &	$\log L_{B}$ & \multicolumn{1}{c}{$F_{5}$} & $\log L_{R}$ & $\log P_{j}$  & $\log P_{j}/L_{d}$ & $\log M_{\mathrm{BH}}$ &\mult{$\log \lambda_{\rm Edd}$} \\
		& & & & & & [ergs$^{-1}$] & \multicolumn{1}{c}{Jy} &[ergs$^{-1}$] & [ergs$^{-1}$] & & [$M_{\odot}$] & \\
		\hline \hline
		0038-0207 &  3C 17            & 0.220 & 18.0 & 0.08 & 0.58 & 43.9 & 2.48000 & 43.2 & 45.7920 &  0.8920 & 8.7 & -1.9  \\
		0044+1211 &  4C +11.06        & 0.226 & 19.0 & 0.26 & 0.28 & 43.8 & 0.22000 & 42.2 & 44.9349 &  0.1349 & 7.8 & -1.1  \\
		0207+2931 &  3C 59            & 0.110 & 16.0 & 0.21 & 0.28 & 44.4 & 0.67000 & 42.0 & 44.7634 & -0.6366 & 8.9 & -1.6  \\
		0224+2750 &  3C 67            & 0.311 & 18.6 & 0.42 & 0.82 & 43.8 & 0.87000 & 43.1 & 45.7063 &  0.9063 & 8.1 & -1.4  \\
		0238-3048 &  IRAS 02366-3101  & 0.062 & 15.0 & 0.22 & 0.30 & 44.2 & 0.00343 & 39.2 & 42.3634 & -2.8366 & 8.6 & -1.5  \\
		0238+0233 &  PKS 0236+02      & 0.207 & 17.7 & 0.11 & 0.46 & 44.1 & 0.12000 & 41.9 & 44.6777 & -0.4223 & 8.8 & -1.8  \\
		0312+3916 &  B2 0309+39       & 0.161 & 18.2 & 0.49 & 0.10 & 44.0 & 0.82200 & 42.5 & 45.1920 &  0.1920 & 8.3 & -1.4  \\
		0342-3703 &  PKS 0340-37      & 0.285 & 18.6 & 0.03 & 0.19 & 44.2 & 0.71000 & 42.9 & 45.5349 &  0.3349 & 8.8 & -1.7  \\
		\hline                                                                                                 
	\end{tabular}            
\end{table}

\begin{table}
	\addtocounter{table}{-1}
	\renewcommand{\thetable}{A\arabic{table}b}
	\small
	\centering                                                                                   
	\caption{The content of the table is analogous to the Table~\ref{tab:4a-BLRG}, but for RLQs instead of BLRGs.}
	\label{tab:4b-4RLQ}
	\begin{tabular}{*{7}{c}*{1}{d{2.3}}*{5}{c}}
		\hline 
		IAU & Name & Redshift & $m_{V}$ & $A_{V}$ & $\kappa_{\star}$ &	$\log L_{B}$ & \multicolumn{1}{c}{$F_{5}$} & $\log L_{R}$ & $\log P_{j}$  & $\log P_{j}/L_{d}$ & $\log M_{\mathrm{BH}}$ &\mult{$\log \lambda_{\rm Edd}$} \\
		& & & & & & [ergs$^{-1}$] & \multicolumn{1}{c}{Jy} &[ergs$^{-1}$] & [ergs$^{-1}$] & & [$M_{\odot}$] & \\
		\hline \hline
		0019+2602 &  4C 25.01      & 0.284 & 15.4 & 0.10 & 0.00 & 45.6 & 0.405 & 42.7 & 45.3634 & -1.2366 & 9.1 & -0.6  \\
		0113+2958 &  B2 0110+29    & 0.363 & 17.0 & 0.21 & 0.00 & 45.2 & 0.311 & 42.8 & 45.4491 & -0.7509 & 9.2 & -1.1  \\
		0157+3154 &  4C 31.06      & 0.373 & 18.0 & 0.18 & 0.11 & 44.8 & 0.394 & 43.0 & 45.6206 & -0.1794 & 9.1 & -1.4  \\
		0202-7620 &  PKS 0202-76   & 0.389 & 16.9 & 0.17 & 0.00 & 45.3 & 0.800 & 43.3 & 45.8777 & -0.4223 & 9.2 & -1.0  \\
		0217+1104 &  PKS 0214+10   & 0.408 & 17.0 & 0.36 & 0.01 & 45.4 & 0.460 & 43.1 & 45.7063 & -0.6937 & 8.9 & -0.7  \\
		0311-7651 &  PKS 0312-77   & 0.225 & 16.1 & 0.32 & 0.00 & 45.2 & 0.590 & 42.6 & 45.2777 & -0.9223 & 8.4 & -0.3  \\
		0418+3801 &  3C 111        & 0.049 & 18.0 & 5.46 & 0.04 & 45.1 & 6.637 & 42.3 & 45.0206 & -1.0794 & 8.8 & -0.8  \\
		0559-5026 &  PKS 0558-504  & 0.138 & 15.0 & 0.15 & 0.00 & 45.1 & 0.121 & 41.5 & 44.3349 & -1.7651 & 7.4 &  0.6  \\
		\hline                                                                                                 
	\end{tabular}       
\end{table}
\end{landscape}

\restoregeometry

%%%%%%%%%%%%%%%%%%%%%%%%%%%%%%%%%%%%%%%%%%%%%%%%%%

\end{document}